\documentclass[a4paper,UKenglish]{lipics-v2021}

\usepackage{listings,multicol}
\usepackage{lstcoq}
\usepackage{amssymb}
\usepackage{subcaption}
\usepackage{semantic}
\usepackage[toc]{appendix}
\usepackage[inline]{enumitem}
\usepackage{hyperref}
\usepackage[capitalise,nameinlink,noabbrev]{cleveref}

\usepackage{xcolor}

\definecolor{ltblue}{rgb}{0,0.4,0.4}
\definecolor{dkblue}{rgb}{0,0.1,0.6}
\definecolor{dkgreen}{rgb}{0,0.5,0}
\definecolor{brightmaroon}{rgb}{0.76, 0.13, 0.28}
\definecolor{burntorange}{rgb}{0.8, 0.33, 0.0}
\definecolor{dkred}{rgb}{0.5,0,0}
\definecolor{bggray}{gray}{0.95}
\definecolor{arsenic}{rgb}{0.23, 0.27, 0.29}

\newcommand{\icode}[1]{\lstinline[mathescape,basicstyle=\small]!#1!}

\newcommand\CIC[0]{CIC}
\newcommand\lambdaCIC[0]{$\lambda_{CIC}$}
\newcommand\CICbox[0]{$\lambda_\square$}
\newcommand\CICboxty[0]{$\lambda_\square^{T}$}

\colorlet{cicterms}{dkblue}
\colorlet{boxtyterms}{dkgreen}
\colorlet{errorcolor}{red}

\newcommand\coqref[2]{\raisebox{-0.3em}{\protect\coqLogo}{\footnotesize\href{https://github.com/AU-COBRA/ConCert/blob/#2}{\texttt{\color{brightmaroon}{#1}}}}}

\newcommand{\coqLogo}{\includegraphics[width=0.75em]{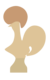}}

\makeatletter
\newtheoremstyle{coqtheorem}
  {}
  {}
  {\itshape}
  {}
  {\bfseries}
  {}
  {\newline}
  {\thmname{#1}\thmnumber{\@ifnotempty{#1}{ }#2}%
   \thmnote{ {\the\thm@notefont(#3).}\newline\extraref}}

\theoremstyle{coqtheorem}
\newtheorem{coqtheoreminner}{Theorem}
\newcommand{\extraref}{}
\newenvironment{coqtheorem}[1]
 {%
  \if\relax\detokenize{#1}\relax
  \else
    \renewcommand{\extraref}{#1}
  \fi
  \begin{coqtheoreminner}}
 {\end{coqtheoreminner}}

\newtheoremstyle{coqdefinition}
  {}
  {}
  {\normalfont}
  {}
  {\bfseries}
  {}
  {\newline}
  {\thmname{#1}\thmnumber{\@ifnotempty{#1}{ }#2}%
   \thmnote{ {\the\thm@notefont(#3).}\newline\extrarefdef}}

\theoremstyle{coqdefinition}
\newtheorem{coqdefinitioninner}{Definition}
\newcommand{\extrarefdef}{}
\newenvironment{coqdefinition}[1]
 {%
  \if\relax\detokenize{#1}\relax
  \else
    \renewcommand{\extrarefdef}{#1}
  \fi
  \begin{coqdefinitioninner}}
 {\end{coqdefinitioninner}}

\usepackage{cleveref}

\title{Formalising Decentralised Exchanges in Coq}

\author{Eske Hoy Nielsen}{Computer Science, Aarhus University}{eske@cs.au.dk}{}{}

\author{Danil Annenkov}{Computer Science, Aarhus University}{danil.v.annenkov@gmail.com}{}{}

\author{Bas Spitters}{Computer Science, Aarhus University}{spitters@cs.au.dk}{}{}

\authorrunning{E. H. Nielsen, D. Annenkov and B. Spitters} 

\Copyright{Eske Hoy Nielsen, Danil Annenkov and Bas Spitters} 



\nolinenumbers 
\hideLIPIcs

\begin{document}

\maketitle

\begin{abstract}
  The number of attacks and accidents leading to significant losses of crypto-assets is growing.
  According to Chainalysis, in 2021, approx.\ \$14 billion has been lost due to various incidents, and this number is dominated by Decentralized Finance (DeFi) applications.
  In order to address these issues, one can use a collection of tools ranging from auditing to formal methods.
  We use formal verification and provide the first formalisation of a DeFi contract in a foundational proof assistant capturing contract interactions.

  We focus on Dexter2, a decentralized, non-custodial exchange for the Tezos network similar to Uniswap on Ethereum.
  The Dexter implementation consists of several smart contracts. This poses unique challenges for formalisation due to the complex contract interactions.
  Our formalisation includes proofs of functional correctness with respect to an informal specification for the contracts involved in Dexter's implementation.
  Moreover, our formalisation is the first to feature proofs of safety properties of the interacting smart contracts of a decentralized exchange.
  We have extracted our contract from Coq into CameLIGO code, so it can be deployed on the Tezos blockchain.

  Uniswap and Dexter are paradigmatic for a collection of similar contracts.
  Our methodology thus allows us to implement and verify DeFi applications featuring similar interaction patterns.
\end{abstract}

\section{Introduction}\label{sec:intro}
Decentralised Finance (DeFi) is an emerging technology that aims to remove third parties in financial transactions.
DeFi crucially utilises distributed ledgers, such as blockchains.
The smart contract technology is built on top of the blockchain and forms a software layer that allows for developing DeFi applications.
One crucial property of smart contracts is \emph{irreversibility}, that is, the execution of a smart contract cannot be simply ``cancelled'' by one of the parties, even though some of the contracts contain errors.

The interest in DeFi is growing, and so are the financial losses due to hacker attacks.
According to Chainalysis~\cite{chainalysis}, approximately \$14 billion was lost in 2021 in cryptocurrency-related crimes, including hacker attacks.
Many of these cases are associated with DeFi applications and quite a few consist of simple programming mistakes.
So, it is important that the software implementing DeFi is bug- and exploit-free.

\emph{Decentralised exchanges} (DEXs) are one of the important applications of the DeFi technology. They enable the exchange of digital assets without involving an intermediary.
Many DEXs are based on \emph{automated market makers} (AMMs) --- autonomous protocols that define prices of the assets involved in the exchanges.
AMMs are implemented as smart contracts deployed on a blockchain; see~\cite{xu2022sok} for an overview.
In their terminology, we are focusing on preventing middleware attacks, which include correct handling of reentrancy, exceptions and mathematical operations.

In this work, we present an implementation methodology for developing and verifying decentralised exchanges in Coq exemplified by the Dexter2 exchange for the Tezos blockchain. Dexter2 is an improved version of the vulnerable Dexter1 protocol. The exchange is based on the Uniswap v1 exchange protocol for Ethereum. 
Our Coq development follows the original Dexter2 implementation written in CameLIGO closely. CameLIGO is a functional programming language for the Tezos blockchain.
A simplified version of this hand-written implementation is deployed and used in the liquidity baking feature of the Tezos Granada protocol.
Hence, the protocol is an integral part of the Tezos blockchain.

The smart contract execution model used by Tezos and a number of modern blockchains\footnote{E.g. Concordium, {\AE}ternity} is based on the message passing/actor model~\cite{Interactions}; this is partially as a reaction to the problems with reentrancy on the Ethereum blockchain.
One of the biggest challenges in the verification of DEXs is the interaction of the smart contracts involved in the DEX, as it not only requires reasoning about state invariants for a single contract but also about inter-contract invariants.
So far, such invariants have not been mechanically verified for this execution model.
The Mi-Cho-Coq framework was used to prove properties for contract functions in isolation, but contract interactions were only subjected to randomised testing\footnote{\url{https://research-development.nomadic-labs.com/follow-up-on-the-verification-of-liquidity-baking-smart-contracts.html}}.
The main Dexter2 contract was also verified for functional correctness by Runtime Verification using the (non-foundational) semi-automatic verifier of the K-framework\footnote{\url{https://runtimeverification.com/blog/dexter-2-s-formal-verification}}.
Safety properties were stated, but not machine-checked.

The Ethereum blockchain uses a different execution model which resembles ordinary procedure calls in imperative programming languages.
This, however, is a deceiving analogy, because any contract being called is potentially adversarial, which opens the door for many potential vulnerabilities.
Most work on the verification of smart contracts focus on single contracts. Recent work on verification~\cite{Bram2021:etherium-vyper} of smart contracts written in Ethereum's Vyper language, a simple language designed with security in mind, proves some properties of interacting contracts.
However, their example of a DEX contract is simpler and does not feature all the interactions considered in our work. Their other examples show that implementing communicating contracts is hard and requires knowledge of common pitfalls of the Ethereum execution model.

\paragraph*{Contributions}
\begin{itemize}
\item We present a methodology of developing DEX applications from implementation in Coq to executable blockchain code.
\item As an example, we provide a full executable implementation of Dexter2 in Coq.
\item We prove functional correctness and important inter-contract invariants of the contracts comprising Dexter2.
\item We extract an implementation in CameLIGO using ConCert's verified extraction framework based on MetaCoq.
\item During the formalisation, we have found minor mismatches between the informal specification and the Dexter2 implementation. Since both were written by expert teams, we believe this merely highlights the fact that specification of programs is best done hand in hand with formalisation.
\end{itemize}

Our contribution is available as part of the ConCert framework (\url{https://github.com/AU-COBRA/ConCert/tree/itp2022}) and consists of approx.\ 5K LOC (implementation, proofs and machinery for reasoning about communicating contracts), plus fixes to CameLIGO extraction and the execution model.
In the text, we refer to our formalisation using the following link format: \coqref{path/to/file.v:lemma\_name}{itp2022/execution/examples/dexter2}.
The path starts at the root of the project's repository, and the optional \texttt{lemma\_name} parameter refers to a definition in the file.

\section{The ConCert Framework}\label{sec:concert}
Before explaining the details of our methodology, we first introduce the ConCert framework, since our work crucially uses its features.
The ConCert framework is implemented in the Coq proof assistant.
The underlying foundation of Coq is the Calculus of Inductive Constructions~\cite{CoquandPaulin:CIC} (\CIC{}) --- a dependently typed functional programming language allowing for expressing both programs and program properties.
Coq also features proof automation through tactics, extension through plugin development and meta-programming through MetaCoq~\cite{MetaCoq}.
Since Coq comes with a functional programming language it is natural to apply it for verification of functional smart contract languages.
It can be done by \emph{embedding} a language to Coq~\cite{Spector-Zabusky:TotalHaskell,ConCert}\footnote{\url{github.com/formal-land/coq-of-ocaml}} or \emph{extracting} a program from Coq formalisation~\cite{letouzey04,CertErasure,ExtractionFromCoqinCoq}.

In the present work, we focus on verification and extraction functionality of the ConCert framework~\cite{ConCert,ConCert-extraction-testing}.
We give a very brief overview of ConCert focusing on the smart contract execution layer, and smart contract extraction.

\subsection{Overview}
\begin{figure}
  \vspace{-1.5em}
  \centering
  \includegraphics[width=12cm]{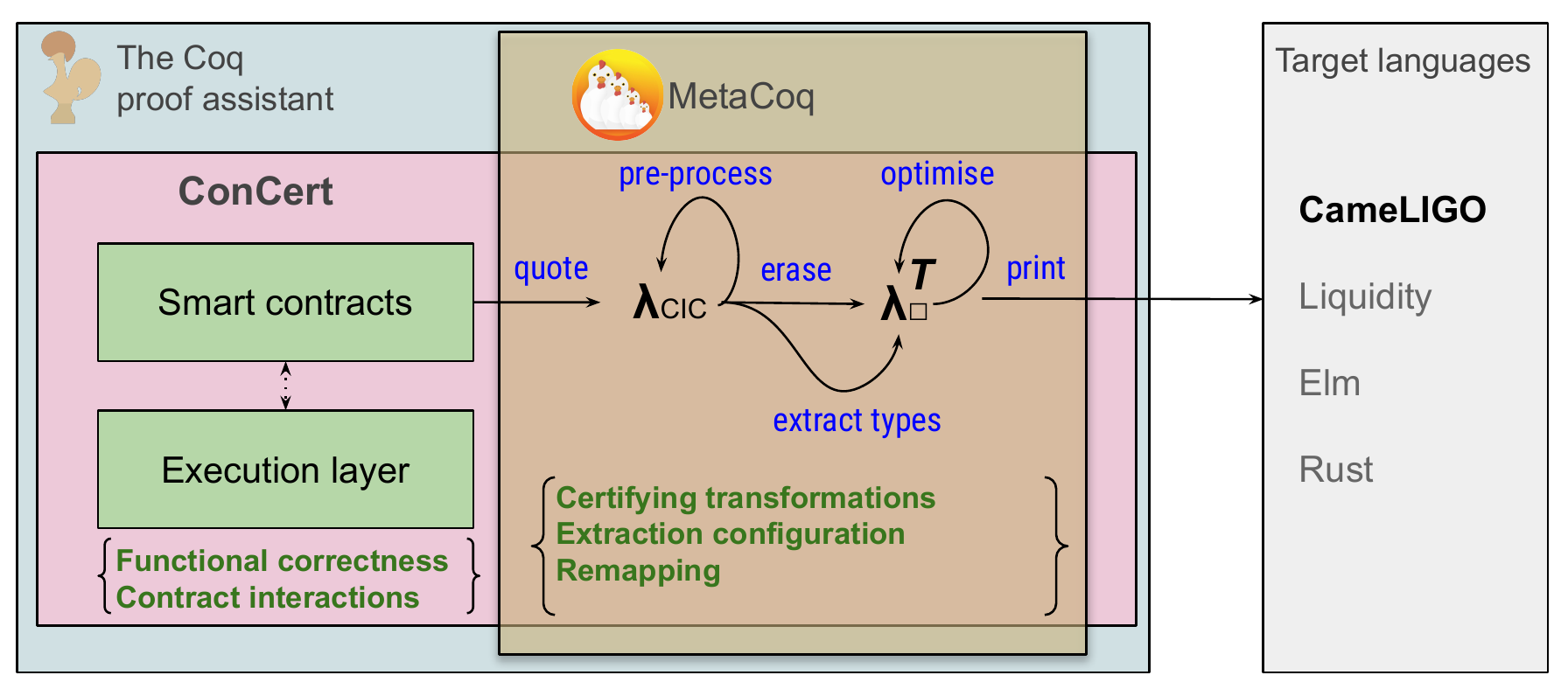}
  \caption{The ConCert framework}\label{fig:concert}
\end{figure}
\noindent%
The overview of the ConCert framework is presented in~\cref{fig:concert}.
We start by developing a smart contract as a function in Coq using the ConCert infrastructure.
For a smart contract, defined as a function, we can state and prove functional correctness properties.
Since smart contracts are just ordinary functions that use blockchain-specific data types defined in ConCert, reasoning about them is as convenient as about any other function in Coq.

One of the key features of the ConCert framework is its support for reasoning about the interaction between several smart contracts and the blockchain.
That is, one can state and prove properties about contract execution traces.
Such proofs crucially use the execution layer to reason about interacting contracts (see more details in~\cref{sec:exec-model}), which enables us to capture properties beyond the mere functional correctness of a single contract invocation (see~\cref{sec:inter-contract}).

One can obtain an executable implementation of the verified code in one of the supported smart contract languages through code extraction.
This code-generation procedure has strong correctness guarantees based on a small trusted computing base of Coq, MetaCoq and the pretty-printers into the target languages.

\subsection{Smart Contract Execution Layer}\label{sec:exec-model}
The execution layer models parts of an account-based blockchain infrastructure for smart contract execution. The account-based model was introduced by Ethereum as an alternative to Bitcoin's UTXO model.
Our model allows us to reason about contract execution traces and hence temporal properties of interacting smart contracts.
Smart contracts in ConCert use a similar model as a number of blockchains where a contract consists of two functions (we use Coq syntax for function signatures):
\begin{lstlisting}[backgroundcolor=\color{white}]
init : Chain -> ContractCallContext -> Setup -> option State
\end{lstlisting}\vspace{-0.5em}
The initialisation function is called after the contract is deployed on the blockchain.
The first parameter of type \icode{Chain} gives access to data about the blockchain (e.g.\ current chain height).
The \icode{ContractCallContext} parameter provides data about the current call (e.g.\ caller address, amount sent to the contract).
\icode{Setup} represents initialisation parameters.
  \begin{lstlisting}[backgroundcolor=\color{white}]
receive : Chain -> ContractCallContext -> State -> option Msg
          -> option (State * list ActionBody)
  \end{lstlisting}\vspace{-0.5em}
The \icode{receive} function represents the main functionality of the contract that is executed for each call to the contract.
\icode{Chain} and \icode{ContractCallContext} are the same as for \icode{init}.
The parameter of type \icode{State} is the current state of the contract; \icode{Msg} is a user-defined type of messages that the contract accepts (the \emph{entrypoints} of the contract).
The result of a successful execution is a new state and a list of \emph{actions} represented with \icode{ActionBody}.
The actions can be transfers, calls to other contracts (including itself), and contract deployments.

Both \icode{receive} and \icode{init} are ordinary Coq functions, making them convenient to reason about.
However, reasoning about the contract functions in isolation is not quite sufficient.
One call to \icode{receive} potentially emits more calls, which can create complex call graphs between deployed contracts.
The execution model is given by an execution trace \icode{ChainedList} --- the reflexive-transitive closure of the proof-relevant \icode{ChainStep} relation, which essentially captures the addition of a single block to the blockchain.
In this step, any actions (like contract calls) are executed.
The type of traces \icode{ChainedList} is defined inductively and thus comes with an induction principle.
ConCert also features a \icode{contract_induction} principle that is more convenient for reasoning for many common properties on execution traces of a single contract.
ConCert provides two runnable implementations of the execution model corresponding to depth-first and breadth-first orders of contract call execution.
For more details about the internals of the execution layer see~\cite{Interactions}.
The Tezos blockchain utilises the depth-first execution order.
However, the ConCert model is not committing to a particular execution order without additional assumptions.
That means that one can obtain stronger results for many properties, namely that a property holds regardless of the execution order of the subsequent calls.

\subsection{Code extraction}\label{sec:extraction-concert}
The Coq proof assistant provides extraction to OCaml, Haskell and Scheme out-of-the-box.
However, it does not support languages for smart contracts.
Moreover, the standard extraction in Coq is implemented in OCaml and is not verified.
Therefore, in order to obtain an implementation suitable for deploying on a particular blockchain, the ConCert Framework provides code extraction.

ConCert's extraction uses the verified erasure procedure of the MetaCoq framework~\cite{MetaCoq} that provides a proof of computational soundness of the important erasure step of extraction.
The erasure procedure takes a Coq term \lambdaCIC{} and replaces computationally-irrelevant parts of a Coq term with a special box node $\square$ and produces an untyped representation \CICbox{}.
On top of this, ConCert adds extraction of typing information leading to the \CICboxty{} representation.
The \CICboxty{} terms are used as an intermediate representation, for which ConCert implements certified optimisations and pretty-printers to several target languages.
ConCert also provides a proof-generating (certifying) pre-processing step featuring inlining and $\eta$-expansion (see~\cite[Section~5.2]{ExtractionFromCoqinCoq}).
In this work, we focus on extraction to CameLIGO --- a functional programming language for the Tezos blockchain. The details of configuring extraction are given in~\cref{sec:extraction}.

\section{The Dexter Exchange}
In order to demonstrate our approach, we implement and verify the Dexter Exchange for the Tezos blockchain.
Dexter features implementations patterns, such as delegating the token bookkeeping functionality to a separate contract, that are common for multi-contract applications.
Moreover, the security audits, together with the various verification and specification projects, make Dexter one of the most carefully studied smart contracts. 
It is thus a good target for formalisation.
In this section, we describe the CameLIGO implementation of the exchange.
We provide the details of how we encode it in ConCert in~\cref{sec:formalisation}.

Dexter is an open-source, non-custodial, decentralized exchange on the Tezos blockchain developed by CamelCase. It allows exchanging tez (the currency of Tezos, also known by its ISO-4217 code XTZ) with FA1.2 tokens. FA1.2 is a fungible token standard for Tezos smart contracts.
In the present work, we focus on Dexter2\footnote{\url{https://gitlab.com/dexter2tz/dexter2tz/-/tree/master/}}, which is a complete rewrite of the original Dexter contract, written in CameLIGO.
It improves upon the original version by
\begin{enumerate*}[label=(\roman*)]
    \item fixing fatal security flaws\footnote{\url{https://research-development.nomadic-labs.com/a-technical-description-of-the-dexter-flaw.html}};
    \item adding a separate liquidity token contract;
    \item adding support for FA2 tokens.
\end{enumerate*}
\begin{figure}
  \vspace{-1.5em}
  \centering
  \includegraphics[width=12cm]{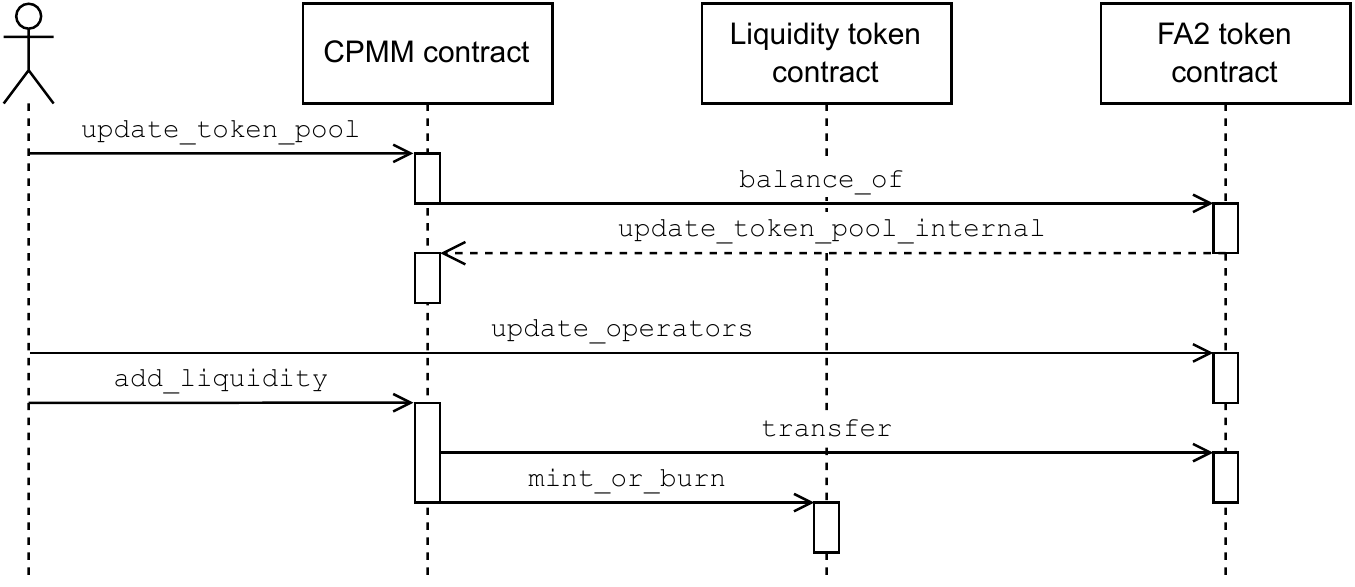}
  \caption{Dexter2 features complex interaction patterns: callbacks, messages to multiple contracts}\label{fig:interaction-patterns}
\end{figure}
FA2 is a newer token contract standard for Tezos. Dexter2 allows trading both FA1.2 and FA2 tokens. Dexter2 delegates the responsibility of tracking ownership of the exchange's assets to a separate liquidity token contract. This means that the Dexter2 exchange consists of three contracts: the main exchange contract with which users interact, the token contract holding the traded tokens, and a liquidity token contract. Dexter2 enables trades directly between two token types, such trades involve two different instances of Dexter2, i.e.\ a total of six different interacting smart contracts. This interaction between multiple contracts makes safety proofs non-trivial. \cref{fig:interaction-patterns} illustrates the interactions between the contracts when a user updates the token pool and adds liquidity.

\subsection{The Dexter2 Smart Contracts}\label{sec:dexter2-smart-contracts}
In this section, we introduce the three contracts in the Dexter2 exchange. The first smart contract is the token contract which implements the token that users can exchange using Dexter2. This contract can be either an FA1.2 or FA2 token contract. In this work, we do not verify the token contract, since Dexter2 is designed to work with many different token contracts. Thus, there is limited gain in verifying a single token implementation.

\subsubsection{Main contract}
The second contract is the \emph{Constant Product Market Maker} contract --- an instance of an automated market maker. This is the main contract, that users use to trade between tokens and tez. It implements a \emph{constant product market} exchange, which is a market where a currency of type $\alpha$ can be traded for a currency of type $\beta$. The market reserves are denoted as $R_\alpha > 0$ and $R_\beta > 0$. The idea of a constant product market exchange is that the value of the market $k=R_\alpha R_\beta$ remains constant when the fee of trading is 0, and can only increase when the fee is positive. Each Dexter2 main contract can only be linked with one token contract.

The main contract has eleven entrypoints. The \icode{addLiquidity} and \icode{removeLiquidity} entrypoints allow liquidity owners to deposit and withdraw from the reserves. Users are required to deposit and withdraw an equivalent value of both currencies. The \icode{xtzToToken, tokenToToken} and \icode{TokenToXtz} entrypoints allow users to trade between tez and tokens at a fixed percentage fee.
The exchange rates are calculated dynamically at each invocation based on current reserves. The main contract tracks the amount of tez, tokens, and liquidity tokens itself, instead of communicating with the other contract, this results in lower gas fees and makes the contract less dependent on execution order.
Because of this, the number of tokens can drift from the actual amount of tokens owned. Therefore, Dexter2 allows updating this counter manually through the \icode{updateTokenPool} entrypoint, which will call \icode{getBalanceOf} on the token contract.
The \icode{setBaker, setManager}, and \icode{setLqtAddress} entrypoints are self-explanatory. A \emph{baker} is the Proof-of-Stake analogue of a miner in a Proof-of-Work protocol. Smart contracts can delegate the power associated to their stake to bakers.
The contract also has a \icode{default} entrypoint that can be used to donate tez to the contract.

\subsubsection{Liquidity token contract}
The third contract is the liquidity token contract. To allow trading, the main contract must have a reasonable reserve of both currencies. To avoid the deployer of the exchange having to put in this large reserve, Dexter2 incentivises users to deposit currency in the reserve. The liquidity token contract is used to track ownership of the reserves.
The liquidity tokens correspond to shares of the main contract's assets.
The liquidity token contract is an FA1.2 token contract with an extra entrypoint for minting and burning tokens. 

Being an instance of the FA1.2 standard, the liquidity token contract includes five entrypoints. 
These entrypoints enable transferring tokens, giving other users access to manage your tokens, and functions for inspecting the contracts state such as \icode{getBalance}. The contract also has a \icode{mintOrBurn} entrypoint which only the main contract may call. This entrypoint is used to mint new liquidity tokens when users deposit assets and burn when they withdraw.

\section{Formalisation}\label{sec:formalisation}
We use the following methodology throughout our development.
First, we implement contracts comprising Dexter2 as Coq functions using the ConCert infrastructure.
For these functions, we verify functional correctness properties (properties of a single function call).
Next, we state and prove single-contract invariants independently.
After that, we connect the individual invariants through a theorem relating incoming and outgoing messages of the interacting contracts to prove inter-contract invariants.
Finally, we use the extraction pipeline of ConCert to obtain a fully functional implementation of the smart contracts.

\subsection{Implementation in ConCert}
To model a smart contract in ConCert, we can either precisely embed it, or consider the extracted smart contract.
For Dexter2, we model the existing contract approximately, but we believe faithfully.
We outline some differences in the sections dedicated to the contracts comprising Dexter2.
We implement the contracts described in~\cref{sec:dexter2-smart-contracts} as Coq functions following the signatures enforced by the ConCert framework, as described in~\cref{sec:exec-model}.
That is, we implement the type of messages (entrypoints), the state type, and the initialisation setup for each contract.
We then extract verified CameLIGO code, which behaves similar to the original contract.
This is currently the most convenient way to obtain fully verified contracts on-chain.

\subsubsection{Liquidity Token Contract}\label{subsec:DexterImplLiquidity}
Coq implementation: \coqref{execution/examples/dexter2/Dexter2FA12.v}{itp2022/execution/examples/dexter2/Dexter2FA12.v}.\\
The entrypoints are modelled using an inductive \icode{Msg} type with a constructor for each entrypoint.
\begin{lstlisting}[language=Coq]
Inductive Msg :=
  | msg_mint_or_burn : mintOrBurn_param -> Msg
  | msg_get_total_supply : getTotalSupply_param -> Msg
  ...
\end{lstlisting}

\noindent The contract's state is represented using record types of Coq.
\begin{lstlisting}[language=Coq]
Record State := build_state { tokens : FMap Address N; allowances : FMap (Address * Address) N;
    admin : Address; total_supply : N}.
\end{lstlisting}
\noindent
Where \icode{FMap} is the type of finite maps (key-value maps), \icode{Address} is part of the ConCert infrastructure, and \icode{N} is the type of binary natural numbers in Coq.
In Coq, one can access the fields of a record using the following projection syntax: \icode{state.(allowances)}.

The liquidity token contract contains several \icode{view} entrypoints, which are entrypoints that return part of the contract's state without modifying the state. However, neither ConCert's execution model nor CameLIGO allow entrypoints to return data. Dexter2 uses callbacks to return the data to the sender. In CameLIGO this is straightforward to implement. However, in ConCert, all entrypoints are bundled together in one inductive definition, which makes modelling callbacks hard since the \icode{view} entrypoint cannot take a name of a particular return entrypoint as input. We get around this problem by defining a \icode{FA12ReceiverMsg} type that contracts must use to be able to receive callbacks. It is defined with one constructor for each callback and a \icode{other_msg : Msg' $\rightarrow$ FA12ReceiverMsg} constructor such that contracts can use it to wrap their existing \icode{Msg} type.
\begin{lstlisting}[language=Coq]
Inductive FA12ReceiverMsg {Msg' : Type} `{Serializable Msg'} :=
  | receive_total_supply : N -> FA12ReceiverMsg
  ...
  | other_msg : Msg' -> FA12ReceiverMsg.
\end{lstlisting}

This approach, however, is not as flexible as in CameLIGO. It enforces the contracts interacting with the Dexter implementation to use particular data types provided by us, while in the original CameLIGO code it is sufficient to have entrypoints with certain names and signatures. However, those restrictions do not affect the safety properties we prove.

Another difference between CameLIGO and the functional language of Coq is that CameLIGO allows throwing errors, i.e.\ stopping executing at any point in the code without returning data. Since functions in Coq are pure, we follow the standard approach and use the \icode{option} monad to faithfully model this behaviour. The only difference is that we do not provide the user with an error message in the case of failure. However, that does not affect the safety of the exchange.

\subsubsection{Main Contract}%
\label{subsec:DexterImplMain}
Coq implementation: \coqref{execution/examples/dexter2/Dexter2CPMM.v}{itp2022/execution/examples/dexter2/Dexter2CPMM.v}.\\
Similarly to the liquidity token implementation, we implement all entry points of the main contract as an inductive type with a constructor for each.
The state is also implemented using records.
\begin{lstlisting}[language=Coq]
Record State := build_state { tokenPool : N; xtzPool : N; lqtTotal : N;
    selfIsUpdatingTokenPool : bool; lqtAddress : Address; ... }.
\end{lstlisting}

The source implementation of the main contract uses compile-time flags to support both FA1.2 and FA2 token standards. However, this means that the compiled code will only support one or the other. In this work, we choose only to model the main contract with FA2 communication.

The main contract needs to be able to receive callbacks from FA2. The FA2 implementation uses callbacks the same way as the liquidity token contract, meaning that the message type for the main contract must be of type \icode{FA2ReceiverMsg}.
\begin{lstlisting}[language=Coq]
Definition Msg := @FA2Token.FA2ReceiverMsg BaseTypes DexterMsg _.
\end{lstlisting}

The \icode{setBaker} entrypoint allows delegating the contract's balance to a baker.
ConCert does not model bakers, so this part of the implementation is omitted. Therefore it is not possible to fully prove functional correctness of this entrypoint.

In~\cref{fig:xtz-to-token} we present an excerpt from the implementation. The implementation uses option monad and custom division and subtraction operations.
CameLIGO, like many other languages, throws runtime errors from division by zero. Moreover, its subtraction operator implicitly converts from natural numbers to integers, such that the subtraction of two natural numbers produces a result of type integer. For other types, subtraction of positive numbers will throw a runtime error if the result is negative. The main contract implementation relies on these runtime errors to ensure the failure of entrypoints on inputs that would break safety.\vspace{-3pt}
\begin{lstlisting}[language=Coq]
Definition sub (n m : N) : option N := do _ <- throwIf (n <? m) ; Some (n $-$ m).
Definition div (n m : N) : option N := do _ <- throwIf (m =? 0) ; Some (n / m).
\end{lstlisting}
\begin{figure}
  \begin{lstlisting}[basicstyle=\scriptsize]
Definition xtz_to_token (chain: Chain) (ctx: ContractCallContext)
                              (state: State) (param: xtz_to_token_param) : result :=
do _ <- throwIf state.(selfIsUpdatingTokenPool) ;
do _ <- throwIf (param.(xtt_deadline) <=? chain.(current_slot))%nat ;
do tokens_bought <- div
  ((amount_to_N ctx.(ctx_amount)) * 997 * state.(tokenPool))
   (state.(xtzPool) * 1000 + ((amount_to_N ctx.(ctx_amount)) * 997)) ;
do _ <- throwIf (tokens_bought <? param.(minTokensBought)) ;
do new_tokenPool <- sub state.(tokenPool) tokens_bought ;
let new_state := state<| xtzPool := state.(xtzPool) + (amount_to_N ctx.(ctx_amount)) |>
                           <| tokenPool := new_tokenPool |> in
let op := token_transfer state ctx.(ctx_contract_address) param.(tokens_to) tokens_bought in
    Some (new_state, [op]).
  \end{lstlisting}
  \caption{\icode{xtz_to_token} implementation}\label{fig:xtz-to-token}
\end{figure}

\subsection{Proof Techniques}\label{subsec:proof-technique}
Previous works on verification of other smart contracts in ConCert have been limited to properties involving incoming or outgoing traffic from arbitrary contracts, and have focused on the perspective of one contract. To fully prove safety of interacting contracts this is not enough. In this section, we discuss how we can achieve proofs of properties reasoning about the interaction between two specific smart contracts and relating their state and traffic. We will limit the scope to only considering invariants over two contract states. However, the methodology should be applicable for reasoning about invariants over any number of contracts.

Typically when proving an invariant over the state, incoming or outgoing actions, one would use the \icode{contract_induction} tactic provided by ConCert. This allows us to induct over chain traces to prove properties that always hold on the contract's state and actions.
However, \icode{contract_induction} does not immediately work when reasoning about interacting contracts. Since ConCert fully formalised the execution model of a blockchain we would be able to prove the property by doing induction over chain traces. However, inducting over chain traces usually gives long and complex proofs even for simple properties, because we have to deal with weak types, (de)serialization, and low-level details of contracts execution.

Here we present an approach to use \icode{contract_induction} to prove smaller isolated invariants over single contracts and combine them to prove invariants over any number of contracts modularly. For contracts A and B we consider proposition \icode{P} of the following form:
\begin{lstlisting}[language=Coq]
P chainHeight currentSlot finalizedHeight
  addrA addrB deployInfoA deployInfoB
  stateA stateB balanceA balanceB
  outActsA outActsB outTxsA outTxsB
  incCallsA incCallsB
\end{lstlisting}
Here \icode{outActsA} are actions emitted by A that have not been executed yet, \icode{outTxsA} are actions emitted by A that have been executed, and \icode{incCallsA} are incoming calls executed by A. 
The general idea is to prove the property from each contract's perspective and then compose the two. Let \icode{R} be a proposition about the first contract and \icode{Q} be a proposition about the second contract. If \icode{R /\\ Q -> P}, then we can use \icode{contract_induction} as we normally would to prove both \icode{R} and \icode{Q}.
We observe that a contract's state is uniquely determined by the initial information given on deployment and all calls executed by the contract, since contracts cannot inspect or modify the state of another contract. That is, any part of a contract's state, outgoing actions, or balance can be expressed as a function \icode{f deployInfo incCalls}. This means we can eliminate both contracts state from \icode{P} and reduce the problem to reasoning about lists of incoming and outgoing messages from the two contracts. To be able to reason about properties on messages sent between the two contracts we need to be able to relate the messages \emph{sent} with the messages \emph{received} by the other contract.

\begin{coqtheorem}{\coqref{execution/theories/InterContractCommunication.v:incomming\_eq\_outgoing}{itp2022/execution/theories/InterContractCommunication.v\#L181}}[Incoming equal to outgoing]\label{theorem:tx-eq-calls}
Let $\pi$ be a reachable blockchain state with contracts $A$ and $B$ deployed.
Let $T_{inB}$ be the set of incoming calls from $A$ to $B$ in the execution trace $\pi$, and let $T_{outA}$ be the set of outgoing calls from $A$ to $B$ in the execution trace $\pi$.
Then $T_{inB} = T_{outA}$.
\end{coqtheorem}

We observe that~\cref{theorem:tx-eq-calls} should hold for any blockchain with message passing execution model.
This allows us to relate the messages sent between the two contracts.
We prove this equality once and for all for any pair of contracts.

\subsection{Properties}
\subsubsection{Functional Correctness}
The developers of Dexter2 provide detailed informal specifications of both the liquidity and main Dexter2 contracts\footnote{\url{https://gitlab.com/dexter2tz/dexter2tz/-/blob/master/docs/informal-spec/dexter2.md}}. We formalise both informal specifications and prove our contract implementations to be functionally correct w.r.t the specification, except for a few exceptions.
Due to the ConCert limitations, we do not prove that the \icode{setBaker} entrypoint emits the correct actions, and do not prove that the contract throws correct error messages.

To prove the implementations correct we prove the following three properties for each entrypoint:
\begin{itemize}
    \item The state is updated correctly
    \item The entrypoint emits the correct operations
    \item Necessary and sufficient conditions for successful execution.
\end{itemize}
In contrast to proving state invariants and inter-contract communication properties proving correctness of entrypoints is straightforward, since the contracts are implemented as functions in Coq.
The proofs are largely automated using custom tactics.

We found several mismatches between the informal specification and the Dexter2 implementation and found minor contradictions in the informal specification.
The flaws in the informal specification (but not the \emph{implementation}) would have led to vulnerabilities similar to those found in Dexter1.
We reached out to the developers who confirmed that the informal specification was incorrect.
Since both the implementation and the specification were written by expert teams, we believe this merely highlights the fact that specification of programs is best done hand in hand with formalisation. However, it is important that informal specification, implementation, and verification are not all done by the same team.

\subsubsection{Inter-Contract Communication}\label{sec:inter-contract}
We demonstrate the inter-contract invariant proof methodology presented in~\cref{subsec:proof-technique} by proving two key safety properties of Dexter2:
\begin{itemize}
    \item Consistency between main contract state and tez balance\footnote{We use tez to denote the native currency of the modelled blockchain as opposed to tokens that can represent various assets}.
    \item Consistency between main contract state and liquidity token contract state.
\end{itemize}
Consistency between the states and balances is critical as exchange rates would be calculated incorrectly otherwise. Related works on Dexter2 formalisation have considered the same safety properties but did not provide machine checked proofs.

\paragraph*{Tez pool correct}
In the Dexter2 exchange, the exchange's tez is held by the main contract. The contract manually tracks its balance of tez in the \icode{xtzPool} variable of its state. To ensure that exchange rates are calculated correctly it is important that this variable reflects the actual amount of tez that the contract owns. Contract state is updated before emitted transfers are executed, thus it is not possible to prove the variable and the actual balance equal at all times. However, it should be the case that they are consistent after each chain of contract calls.

\begin{coqtheorem}{\coqref{execution/examples/dexter2/Dexter2CPMM.v:contract\_balance\_correct}{itp2022/execution/examples/dexter2/Dexter2CPMM.v\#L1854}}[Tez pool correct]\label{theorem:xtz-correct}
 Let $\pi$ be a reachable blockchain state with the main contract deployed. Let $b$ be the contract balance, and let $T$ be the set of transactions from the main contract that is in the execution queue of $\pi$. Let $x$ be the initial amount of tez given to the contract during deployment. If $x = 0$ and $T = \emptyset$ (no pending transfers) then \icode{xtzPool} is equal to the contract balance $b$.%
\footnote{Note that we do not assume explicitly that the initial value of \icode{xtzPool} is zero.
In ConcCert, the \icode{init} function takes care of generating valid initial state with zero \icode{xtzPool}.
The extracted code features a wrapper that allows for initialising with valid data.}
\end{coqtheorem}
In Coq, we first prove a more general version of the theorem stating that \icode{xtzPool} is equal to the contract balance minus any pending transfers. This is proven by induction on execution traces. From this~\cref{theorem:xtz-correct} follows trivially. Using the theorem we can further prove that Dexter2 never attempts to transfer more tez than it has.

\paragraph*{Liquidity supply correct}
Liquidity tokens are stored in the liquidity token contract of the Dexter2 exchange.
However, the main contract needs to know the total amount of liquidity tokens when calculating exchange rates.
To decrease gas fees the main contract tracks this number itself instead of requesting it from the liquidity token contract.
Therefore, it is critical that the counters in the two states are consistent.
Note also that the pattern of delegating token management to a separate token contract appears in other settings (cf.\ \texttt{auction\_token} example in~\cite{Bram2021:etherium-vyper}).
The same implementation and proof techniques are applicable for such contracts as well.

We would like to formulate the property \emph{modularly}, meaning that it should hold for \emph{any} liquidity token \emph{satisfying certain interface}.
The fact that Coq is based on dependent types allows us to define the token interface along with a property, that it must satisfy in order for the inter-contract invariant to hold.
We define the required property for any given liquidity token contract $\mathcal{L}$ with compatible entry points and state as follows:
\begin{coqdefinition}{\coqref{execution/examples/dexter2/Dexter2FA12.v:LqtTokenInterface}{itp2022/execution/examples/dexter2/Dexter2FA12.v\#L131}}[Liquidity token condition]\label{def:lqt-interface}
  For any $\pi$ reachable blockchain state with $\mathcal{L}$ deployed at address $a_L$.
  Let $a_M$ be the address of the main contract recorded in the initial state of $\mathcal{L}$.
  Let $C_M$ be the set of executed calls from $a_M$ to $a_L$ in the execution trace $\pi$, let $l_L$ be the number of liquidity tokens, and let $i_L$ be the initial amount of liquidity tokens.
  Then $l_L = i_L + sum (map\ mintedAndBurnedTokens\ C_M)$, that is the number of tokens is the initial supply plus minted tokens minus burned tokens.
\end{coqdefinition}

Now, given any implementation of the liquidity token that satisfies~\cref{def:lqt-interface}, we prove the liquidity supply invariant in~\cref{theorem:liquidity-correct}.

\begin{coqtheorem}{\coqref{execution/examples/dexter2/Dexter2CPMM.v:lqt\_pool\_correct\_interface}{itp2022/execution/examples/dexter2/Dexter2CPMM.v\#L2889}}[Liquidity supply correct]\label{theorem:liquidity-correct}
  Let $\pi$ be a reachable blockchain state with the main contract deployed at address $a_M$ and any compatible liquidity token contract that satisfies~\cref{def:lqt-interface} deployed at address $a_L$.
  Let $T$ be the set of transactions from $a_M$ to $a_L$ in the execution queue of $\pi$.
  Let $l_M$ be the liquidity token counter in the main contracts' state and let $l_L$ be the actual amount of liquidity tokens in the liquidity token contract.
  Let $i_M$ and $i_L$ be the initial amounts of liquidity tokens in the two contracts after deployment.
  If the contracts are correctly paired, $i_M = i_L$, and $T = \emptyset$ then the contracts have consistent state $l_M = l_L$.
\end{coqtheorem}
In Coq, the invariant is formulated the following way:
\begin{lstlisting}[basicstyle=\scriptsize,language=Coq]
Definition lqtTotal_total_supply_invariant
  (* Given a liquidity token interface *)
  (i_lqt_contract : LqtTokenInterface) :=
   forall bstate caddr_main caddr_lqt (trace : ChainTrace empty_state bstate),
      (* main contract deployed at $a_M$ *)
      env_contracts bstate caddr_main = Some (contract : WeakContract) ->
      (* liquidity token deployed at $a_L$ *)
      env_contracts bstate caddr_lqt = Some (i_lqt_contract.(lqt_contract) : WeakContract) -> 
      exists state_main state_lqt depinfo_main depinfo_lqt,
        contract_state bstate caddr_main = Some state_main /\
        contract_state bstate caddr_lqt = Some state_lqt /\
        deployment_info Setup trace caddr_main = Some depinfo_main /\
        deployment_info Dexter2FA12.Setup trace caddr_lqt = Some depinfo_lqt /\
        (* initial amount $i_M$ *)
        let initial_tokens_main := lqtTotal_ (deployment_setup depinfo_main) in
        (* initial amount $i_L$*)
        let initial_tokens_lqt := initial_pool (deployment_setup depinfo_lqt) in
        (state_main.(lqtAddress) = caddr_lqt ->
        state_lqt.(admin) = caddr_main ->
        (* $i_M = i_L$*)
        initial_tokens_main = initial_tokens_lqt ->
        (* set of transactions from $a_M$ to $a_L$ $T = \emptyset$*)
        filter (actTo state_main.(lqtAddress)) (outgoing_acts bstate caddr_main) = [] ->
        (* $l_M$ = $l_L$ *)
        state_main.(lqtTotal) = state_lqt.(total_supply)).
\end{lstlisting}

We prove \cref{theorem:liquidity-correct} using the method described in \cref{subsec:proof-technique}.
We first prove a single-contract invariant describing the main contract's liquidity counter in terms of the contract's incoming and outgoing messages.
\begin{coqtheorem}{\coqref{execution/examples/dexter2/Dexter2CPMM.v:lqt\_total\_correct}{itp2022/execution/examples/dexter2/Dexter2CPMM.v\#L2792}}[Main contract liquidity counter correct]\label{lqt-main-correct}
Let $\pi$ be a reachable blockchain state with the main contract deployed at address $a_M$. Let $a_L$ be the address of the liquidity token contract that the main contract is paired with.
Let $T_L$ be the set of executed transactions from $a_M$ to $a_L$ in the execution trace $\pi$, and let $T$ be the set of transactions from $a_M$ to $a_L$ in the execution queue of $\pi$.
Let $l_M$ be the main contract's liquidity token counter and let $i_M$ be the initial amount of liquidity tokens.
Then $l_M = i_M + sum (map\ mintedAndBurnedTokens\ T_L) + sum (map\ mintedAndBurnedTokens\ T)$
\end{coqtheorem}

Now, the proof \cref{theorem:liquidity-correct} follows from using \cref{theorem:tx-eq-calls} that connects the two single-contract invariants: \cref{lqt-main-correct} and the liquidity token condition.
Having \cref{theorem:liquidity-correct}, we can separate liquidity token implementation from the main contract.
In our development, we show that the liquidity token described in~\cref{subsec:DexterImplLiquidity} indeed satisfies the liquidity token condition.
The statement that connects all implementation pieces together is given by\\
\coqref{execution/examples/dexter2/Dexter2CPMM.v:lqt\_pool\_correct\_lqt\_fa12}{itp2022/execution/examples/dexter2/Dexter2CPMM.v\#L2922}

\subsection{Extraction}\label{sec:extraction}
\begin{figure}
  \begin{subfigure}[t]{.5\textwidth}
      \begin{lstlisting}[basicstyle=\scriptsize,caption=Remapping]
Definition TT_remap_arith :=
[   remap <%% Z %%> "int"
  ; remap <%% N %%> "nat"

  ; remap <%% N.add %%> "addN"
  ; remap <%% N.sub %%> "subNTruncated"
  ; remap <%% N.modulo %%> "moduloN"
  ...
  ; remap <%% Z.add %%> "addInt"
  ; remap <%% Z.sub %%> "subInt"
  ; remap <%% div %%> "divN_opt"
  ...
  (* conversion [N] -> [Z] *)
  ; remap <%% Z.of_N %%> "z_of_N"
  ...
].
      \end{lstlisting}
  \end{subfigure}
  \begin{subfigure}[t]{.5\textwidth}
    \begin{lstlisting}[basicstyle=\scriptsize,language=Caml,caption=CameLIGO prelude]
[@inline] let addN (a: nat ) (b: nat ) = a + b

let subNTruncated (n: nat) (m: nat) : nat =
  if n < m then 0n else abs (n$-$m)

let moduloN (n: nat) (m: nat) : nat =
  match ediv n m with
  | Some (_,r) -> r
  | None -> 0n

let divN_opt (n: nat) (m: nat) : nat option =
  match ediv n m with
  | Some (q,_) -> Some q
  | None -> None

let z_to_N (i: int) : nat = if i < 0 then 0n else abs i
      \end{lstlisting}
  \end{subfigure}\vspace{-12pt}
  \caption{Remapping definition for extraction}\label{fig:extraction-config}
\end{figure}

In order to produce an executable implementation that can be deployed on an actual blockchain, one needs to generate smart contract code.
A well-known technique called \emph{code extraction} can be used for that purpose.
However, the standard code extraction targets conventional (mostly functional) programming languages and not the languages for smart contracts.

We utilise the extraction pipeline of ConCert (\cref{sec:concert}) in order to obtain source code for the contracts comprising Dexter2 in the Coq proof assistant.
We choose to use CameLIGO as a target language since it is supported by the Tezos blockchain.
CameLIGO is a functional smart contract language from LIGO family of languages used as high-level user-friendly programming languages for the Tezos blockchain.
CameLIGO uses the syntax of OCaml, but has some restrictions with respect to data types and recursive functions; see \cite[Section~5.3]{ExtractionFromCoqinCoq}.
Code extraction to smart contract languages is somewhat different from extraction to conventional languages.
Smart contracts utilise the blockchain infrastructure that is accessible as part of their runtime.
Therefore, some definitions from the execution layer of ConCert must be remapped to their counterparts in the target blockchain.
Moreover, one has to carefully remap arithmetic operations to primitives of a target language since extracting Coq representations would be either inefficient or impossible due to restrictions on data types.

We summarise our extraction effort in the following points.
\begin{itemize}
\item We update CameLIGO pretty-printing to support polymorphic definitions.
  This feature was added recently and was not supported in ConCert before.
\item We specify a translation table that remaps Coq arithmetic operations for binary integer numbers \icode{Z} and binary natural numbers \icode{N}, operations on finite maps provided by the \texttt{std++} library, and blockchain-specific infrastructure to CameLIGO.
\item We extract the main contract and the liquidity token contract of Dexter2 to CameLIGO code.
\end{itemize}

The important feature of the extraction functionality provided by ConCert is the ability to run the whole extraction pipeline inside Coq.
This feature allows us to keep the trusted computing base smaller (e.g.\ the standard extraction is written in OCaml and executed outside of the proof assistant).

\cref{fig:extraction-config} presents a fragment of the remapping for numeric data types and numerical operations along with corresponding hand-written CameLIGO definitions that we call the \emph{prelude}.
We use \icode{remap <\%\% coq_def \%\%> "cameligo_def"} to produce an entry in the translation table, where the \icode{<\%\% coq_def \%\%>} notation uses MetaCoq metaprogramming facilities to resolve the full name of the given definition.
Note that remapping enhances the trusted computing base.
Therefore, it has to be done clearly and should be easy to audit.
To have better control over how the primitives are remapped, we provide a prelude where we specify explicitly arithmetic operations for each numeric data type.
The prelude serves as a model for Coq arithmetic in CameLIGO.
Many operations are simple wrappers around overloaded \icode{+},\icode{*}, etc. of CameLIGO.
Even though it looks trivial, there are subtle discrepancies between some operations in Coq in CameLIGO.
For example, we need to take special care of subtraction and division.
Subtraction of two natural numbers \icode{nat} in CameLIGO is not a natural number, but an integer \icode{int}.
In Coq, the subtraction of natural numbers is truncated, that is, \icode{n - m} returns zero if \icode{n < m}.
We implement \icode{subNTruncated} that has the same behaviour as in Coq: it returns a natural number and it is truncated in the same way.
We also use safe partial division function \icode{div : N -> N -> option N}.
Under the hood, it uses Euclidean division from the standard library of Coq.
However, the standard division in Coq returns zero when the divisor is zero.
We therefore implement a partial division function \icode{divN_opt} in CameLIGO and remap our \icode{div} to it.
Note that the CameLIGO division is also implemented as Euclidean division followed by projection of the first component (the quotient), thus there is no semantic mismatch between the two implementations. The \icode{moduloN} function is implemented similarly, projecting the second component.

Parts of the blockchain infrastructure, such as call context, is created using Tezos build-in operations for accessing meta-information.
That is, we create an instance of the record type \icode{cctx} with fields corresponding to ConCert's \icode{ContracCallContext}.
\begin{lstlisting}[keepspaces=true]
let cctx_instance : cctx = { ctx_origin_ = Tezos.source;
                                   ctx_from_ = Tezos.sender;
                                   ctx_contract_address_ = Tezos.self_address;
                                   ctx_contract_balance_ = Tezos.balance;
                                   ctx_amount_ = Tezos.amount }.
\end{lstlisting}
This instance is passed as a parameter to the extracted contract in the wrapping code generated by ConCert.

Contract calls are remapped to idiomatic CameLIGO code snippets that uses \icode{Tezos.get_contract_opt} for resolving a given address to a contract instance.
Note, however, that such resolution might fail, but it would correspond to a failed contract call 

\begin{figure}
  \begin{lstlisting}[language=Caml,basicstyle=\scriptsize]
let xtz_to_token (chain: chain) (ctx: cctx) (state: dexter2CPMM_State) (param: dexter2CPMM_xtz_to_token_param) : dEX2Extract_result = 
  match throwIf state.selfIsUpdatingTokenPool with 
    Some val0 -> 
      (match throwIf (lebN param.xtt_deadline (current_slot chain)) with 
        Some val1 -> 
        (match (divN_opt (multN (multN (mutez_to_natural (ctx_amount ctx)) 997n) state.tokenPool)
                (addN (multN state.xtzPool 1000n) (multN (mutez_to_natural (ctx_amount ctx)) 997n))) with 
          Some val2 -> 
          (match throwIf (ltbN val2 param.minTokensBought) with 
            Some val3 -> 
            (match sub state.tokenPool val2 with 
              Some val4 -> 
              let new_state = set_State_tokenPool (fun (a : nat) -> val4)
                (set_State_xtzPool (fun (a : nat) -> addN state.xtzPool (mutez_to_natural (ctx_amount ctx))) state) in 
              let op = token_transfer state (ctx_contract_address ctx) param.tokens_to val2 in 
                Some (new_state, (op :: ([]: (operation) list)))
            | None  -> None: (dexter2CPMM_State *  (operation) list) option)
         ...    (* similar "None" cases are omitted *)
  | None  -> (None: (dexter2CPMM_State *  (operation) list) option)
  \end{lstlisting}\vspace{-5pt}
  \caption{Extracted CameLIGO code}\label{fig:extracted-code}
\end{figure}

In \cref{fig:extracted-code} we present an excerpt from the extracted code, namely code for \icode{xtz_to_token} (see Coq code in \cref{fig:xtz-to-token}).
All arithmetic operations are remapped to the ones we specified in the prelude that give us a way of auditing the remapping.
One difference with the handwritten version is multiple nested pattern-matchings.
CameLIGO supports an effectful construct \icode{failwith}, so the handwritten implementation uses it to return earlier from the function when validation fails.
In Coq code, we use the \icode{option} monad and \icode{do}-notations similar to Haskell giving a similar user experience.
After unfolding the monadic operations we get nested pattern-matching in the extracted code.
Note that all parametric type constructors not carrying values (e.g.\ \icode{None} and empty lists) require explicit type annotations.
In this case, typing information preserved by the ConCert extraction pipeline allows us to produce well-typed code.
The functions with names \icode{set_State_*} originate from the record update syntax (also part on ConCert) that generates setters to record fields automatically.

In general, the proof-producing inlining step before extraction allows us to remove type class instances from the extracted code.
The \icode{option} monad, that we mentioned before, adds a lot of convenience along with the \icode{do}-notation.
However, it relies on type classes, which are not supported in CameLIGO.
Therefore, we use inlining in order to unfold the type class instances and reveal their implementation.

As a result of extraction, we obtain two source files corresponding to the Dexter2 contracts described in~\cref{subsec:DexterImplLiquidity,subsec:DexterImplMain}.
The source files contain CameLIGO code that can then be compiled to Michelson as part of the build process.
The Michelson code can be deployed on the Tezos blockchain.

\section{Related Work}

Several works have already explored the correctness and safety of the Dexter2 decentralized exchange. Dexter2 was formalised and verified for functional correctness by Nomadic Labs\footnote{\url{https://gitlab.com/nomadic-labs/mi-cho-coq/-/blob/dexter_fa12lqt-verification/src/contracts_coq/dexter_spec.v}} using Mi-Cho-Coq. Mi-Cho-Coq~\cite{Mi-Cho-Coq} is a Coq framework for formal verification of Michelson smart contracts. Michelson is the low-level on-chain language that CameLIGO compiles to. The main Dexter2 contract was also verified for functional correctness and safety by Runtime Verification\footnote{\url{https://runtimeverification.com/blog/dexter-2-s-formal-verification}} using the semi-automatic verifier of the K-framework~\cite{10.1145/2983990.2984027}. However, their safety properties are not machine-checked and relied on assertions on the behaviour of the other contracts used in the exchange. Later work by Nomadic Labs has verified some safety properties of the main contract\footnote{\url{https://research-development.nomadic-labs.com/progress-report-on-the-verification-of-liquidity-baking-smart-contracts.html}} and performed property-based testing to test safety properties that could not be proven due to limitations of Mi-Cho-Coq\footnote{\url{https://research-development.nomadic-labs.com/follow-up-on-the-verification-of-liquidity-baking-smart-contracts.html}}.

Another line of work is the verification of interacting smart contracts in general.
VerX~\cite{Permenev2020:VerX} is an automatic verifier for temporal properties of interacting contracts on Ethereum. It is thus similar to ConCert, which also considers temporal properties.
VerX uses temporal quantifiers to express properties of interacting contracts, which are then translated to predicates on execution traces. However, they do not seem to prove complex properties of interacting contracts.


The authors of~\cite{Bram2021:etherium-vyper} verify smart contracts written in the Vyper language for the Ethereum blockchain.
The work uses a flavour of separation logic for a core of Vyper. The tool facilitates modular proofs about inter-contract invariants similar to those presented in~\cref{sec:inter-contract}.
The technique is implemented as an automated verification tool \texttt{2vyper} using the Z3 SMT-solver.
The work verifies some properties of the Uniswap\footnote{\url{https://docs.uniswap.org/protocol/V1/introduction}} contract.
Uniswap has similar functionality to Dexter2, but the implementation verified in \texttt{2vyper} has the main contract and the liquidity token merged to a single smart contract.
This facilitates verification, e.g.\ the invariant that the total number of tokens is equal to the sum of balances, becomes a single-contract invariant.
The paper features an example of the auction contract that uses a separate token contract to handle transfers along with an inter-contract invariant.
It would be interesting to see whether this functionality is sufficient to implement a version of Uniswap that is split into three contracts, similarly to Dexter in the present paper.
Note, however, that Ethereum's execution model is different from the one of ConCert and Tezos.
Therefore, the implementation and the statements of invariants are not directly comparable.
From the implementation of the auction contract, it is clear that implementing interacting contracts for Ethereum is hard and requires knowledge of common pitfalls and programming patterns, such as using lock variables.

\section{Conclusions}
We have presented a methodology for developing and verifying decentralised exchanges.
We have used the Dexter2 exchange as an example, which we implemented in Coq using the ConCert framework.
Dexter2 is paradigmatic and exhibits many common implementation idioms used in blockchains based on the message passing/actor model.
That demonstrates that our methodology is suitable for verifying implementations of similar DEX applications.
A key feature of our approach is the ability to reason about contract interactions and inter-contract invariants.
Another important functionality is code extraction of ConCert that allows us to generate smart contract code.
As a result, one gets a fully functional version of the exchange in CameLIGO smart contract language for the Tezos blockchain.

We believe that most of our methodology is not specific to Coq and can be implemented in other proof assistants based on type theory.
Our verified extraction of smart contracts depends on well-developed meta-programming, which may not be readily available in all proof assistants.

In the future, we would like to connect our Dexter2 development to our ongoing work on formalisation of token standards.

\vspace{-5pt}
\section*{Acknowledgements}%
\vspace{-5pt}
We would like to thank the LIGO team and in particular Tom Jack, Rapha\"el Cauderlier, Exequiel Rivas, R\'emi Les\'en\'echal and Gabriel Alfour for the pleasant discussions.
This research was partially supported by a grant from Nomadic Labs and by the Concordium Blockchain Research Center.

\bibliography{paper}

\end{document}